# Chapter 17

# Semiclassical Methods of Deformation Quantisation in Transport Theory

M. I. Krivoruchenko

*Institute for Theoretical and Experimental Physics*
*B. Cheremushkinskaya 25, 117259 Moscow*
*Russia*

## 1. Introduction

Among the equivalent formulations of quantum mechanics, Heisenberg's matrix mechanics and quantum theory, based on the Schrödinger wave equation, are the most technically advanced. Since the advent of quantum mechanics, these two schemes have provided the mathematical tools and the primary basis for the description of quantum phenomena. The path integral method developed by Dirac and Feynman has important advantages in the quantisation of gauge theories. We discuss here the formulation of quantum mechanics in phase space, known as deformation quantisation.

Deformation quantisation uses the Wigner-Weyl association rule (Weyl, 1927, 1929, 1931; Wigner, 1932) to establish a one-to-one correspondence between the functions in the phase space and the operators in the Hilbert space. Wigner's function appears as the Weyl symbol of the density matrix. A consistent dynamical description of the systems with the help of the Wigner function leads to deformation quantization. A useful formulation of the Wigner-Weyl association rule was proposed by Groenewold (1946) and Stratonovich (1957).

Groenewold introduced a non-commutative associative ⋆-product (star-product) of the functions in the phase space (Groenewold, 1946). The evolution of the quantum systems is determined by the antisymmetric part of the ⋆-product (Groenewold, 1946; Moyal, 1949), known as the Moyal bracket. The Moyal bracket represents the quantum deformation of the Poisson bracket. Deformation quantisation preserves many features of classical Hamiltonian dynamics.

The formulation of deformation quantisation is based on the Wigner function and the Moyal bracket (i.e., the ⋆-product). The Wigner-Weyl association rule is necessary to prove the equivalence of deformation quantisation and the standard formalisms of quantum mechanics.

Extensive literature has reported on the formulation of quantum mechanics in the phase space and the ⋆-product. We refer the reader to excellent reviews by Bayen et al. (1978a, 1978b), Carruthers & Zachariasen (1983), Balazs & Jennings (1984), Hillery et al. (1983), Karasev & Maslov (1991), and Osborn & Molzahn (1995), where one may find additional references. Wigner's function, as a fundamental object of deformation quantisation, has numerous applications in many-body physics, kinetic theory, collision theory, and quantum chemistry. Transport models, originally created to simulate chemical reactions, have been modified and are widely used to describe heavy-ion collisions.

Deformation quantisation does not have many recognised successful applications in quantum theory. Recently, attempts have been made to use specific properties of the formalism to investigate semiclassical expansion (Osborn & Molzahn, 1995; McQuarrie et al., 1998) and to calculate the determinants of one-loop operators in field theory (Pletnev & Banin, 1999; Banin et al., 2001) and the high-order corrections to the Bohr-Sommerfeld quantisation rule (Gracia-Saz, 2004; Cargo et al., 2005). Potential applications of deformation quantisation in transport models are presented in this review.

Transport models for heavy-ion physics are designed for phenomenological descriptions of the complex dynamics of nuclear collisions. Several types of advanced transport models are based on the Boltzmann-Uhlenbeck-Ueling equations (BUU) (Blaettel et al., 1993), (relativistic) quantum molecular dynamics (QMD/RQMD) (Sorge et al., 1989; Aichelin, 1991; Faessler, 1992), or antisymmetrised molecular dynamics (AQMD) (Feldmeier & Schnack, 1997). These approaches have the correct classical limit and they contain special plug-ins and quantum-mechanical attributes, such as Pauli blocking for binary collisions of fermions. Numerical solutions are implemented through the distribution of test particles (BUU) or centroids of wave packets (QMD, AQMD) for classical trajectories in the phase space. For AQMD, the wave packets are antisymmetrised in their parameters. The transport models provide a solid basis for a phenomenological description of a variety of complex nuclear collisions phenomena. However, quantum coherence effects and non-localities are beyond the scope of these models. The internal consistencies of the approximations in the models remain a subject of debate, promoting further developments (see, for example, papers by Kohler (1955) and Feldmeier & Schnack (1997) and references therein).

The most striking feature of the transport models is the depiction of the trajectories in the phase space, which are test particles or centroids of wave packets. The evolution of the system of classical particles can be calculated using standard programs to solve first-order ordinary differential equations (ODEs). At the same time, the evolution of wave functions of many-body systems is a field-theoretic problem with an infinite number of degrees of freedom that cannot be solved either analytically or numerically.

Any simulation of many-body quantum dynamics must be based on a concept of trajectories, which is the only attribute allowing access to an approximate description of complex quantum systems.

The concept of phase-space trajectories arises naturally in the formalism of deformation quantisation through the Wigner transformation of the operators of the canonical coordinates and momenta in the Heisenberg representation. These trajectories satisfy the quantum version of Hamilton's equations (Osborn & Molzahn, 1995; Krivoruchenko & Faessler, 2006b) and are the characteristics by which the time-dependent Weyl's symbols for the other operators can be determined (Krivoruchenko & Faessler, 2006b; Krivoruchenko et al., 2006c, 2007). In the classical limit, quantum characteristics reduce to classical trajectories.

Knowledge of the quantum phase flow, i.e., the quantum trajectories, is equivalent to a complete knowledge of the quantum dynamics.

In this Chapter, we provide an introduction to deformation quantisation and demonstrate the usefulness of the formalism in solving the evolution problem for many-body systems in terms of semiclassical expansion. We show that, in any fixed order of expansion over the Planck's constant, the evolution problem can be reduced to a statistical-mechanics problem of calculating an ensemble of quantum characteristics in the phase space and their Jacobi fields. In comparison with the corresponding rules of classical statistical mechanics, the rules for computing the probabilities and time-dependent averages of observables are modified. The evolution equations represent a finite system of first-order ODEs for quantum trajectories in the phase space and the associated Jacobi fields (Krivoruchenko et al., 2006c). The method of quantum characteristics allows the consistent inclusion of specific quantum effects, such as non-localities and coherence, in the description of the propagation of particles in the transport models.

In the next section, the Wigner-Weyl association rule is described, and the concept of the ★-product is introduced using the Groenewold method (Groenewold, 1946). Section 3 is devoted to the properties of the quantum characteristics. We explore the transformation properties of the canonical variables and functions in phase space under unitary transformations in Hilbert space using the Wigner-Weyl correspondence rule. The role of quantum characteristics coincides with the role of characteristics in a solution of the classical Liouville equation. Section 4 starts from the semiclassical expansion of ★-functions around the normal functions. The results are then applied to the decomposition of the functions of the quantum characteristics. Quantum characteristics, when expanded in a power series of Planck's constant, are found by solving a coupled system of ODEs for quantum characteristics and the associated Jacobi fields. The numerical methods for solving many-body scattering problem, including the rule for the calculation of the average values of physical observables, are discussed in Section 5. Special features of the scattering problem are discussed in Section 6.

In the field of deformation quantisation, the terminology is not well established. Deformation quantisation is synonymous with Weyl-Groenewold quantisation and ★-product quantisation. The ★-product is synonymous with the Moyal product. Wigner's image of an operator is Weyl's symbol. The Moyal bracket is also known as the sine bracket. The quantum Liouville equation is synonymous with the Groenewold equation and is the Wigner image of the von Neumann equation.

## 2. The Wigner-Weyl association rule, the ★-product and the Wigner function

In Hamiltonian formalism, classical systems with $n$ degrees of freedom are described by $2n$ canonical coordinates and momenta

$$\xi^i = (q^1,...,q^n,p_1,...,p_n) \in \mathbb{R}^{2n}.$$

The Poisson bracket for these variables takes the simple form

$$\{\xi^k, \xi^l\} = -I^{kl}. \tag{1}$$

The matrix

$$\|I\| = \begin{Vmatrix} 0 & -E_n \\ E_n & 0 \end{Vmatrix},$$

where $E_n$ is the identity matrix, endows the phase space with a symplectic structure. In the following, we use the form $I^{ij}$ to lower and raise the indices, e.g., $A^i = A_j I^{ji}$, $A_i = I_{ij} A^j$, where $I_{ij} = I^{ji}$ and $A$ is a vector in the phase space.

In quantum mechanics, the canonical variables are mapped into operators of the canonical coordinates and momenta in a Hilbert space:

$$\mathfrak{x}^i = (\mathfrak{q}^1,...,\mathfrak{q}^n,\mathfrak{p}_1,...,\mathfrak{p}_n) \in Op(L^2(\mathbb{R}^n)).$$

These operators obey the commutation relations

$$[\mathfrak{x}^k, \mathfrak{x}^l] = -i\hbar I^{kl}. \tag{2}$$

The operators $\mathfrak{f} \in Op(L^2(\mathbb{R}^n))$ are denoted by Gothic letters, and the functions in phase space $\mathbb{R}^{2n}$ are denoted by Latin letters.

The Wigner-Weyl correspondence $\xi^i \leftrightarrow \mathfrak{x}^i$ extends to arbitrary functions and operators. A set of operators $\mathfrak{f} \in Op(L^2(\mathbb{R}^n))$ in the Hilbert space is a set closed under multiplication by $c$-numbers and summation. This set forms a vector space $\mathbb{V}$. The elements of its basis can be numbered by the phase space coordinates $\xi^i \in \mathbb{R}^{2n}$. Typically, the Weyl-Groenewold basis is used:

$$\mathfrak{B}(\xi) = (2\pi\hbar)^n \delta^{2n}(\xi - \mathfrak{x}) = \int \frac{d^{2n}\eta}{(2\pi\hbar)^n} \exp(-\frac{i}{\hbar} \eta_k (\xi - \mathfrak{x})^k). \tag{3}$$

The vectors $\mathfrak{B}(\xi) \in \mathbb{V}$ satisfy the following properties:

$$\mathfrak{B}(\xi)^+ = \mathfrak{B}(\xi),$$
$$\int \frac{d^{2n}\xi}{(2\pi\hbar)^n} \mathfrak{B}(\xi) = 1,$$
$$\int \frac{d^{2n}\xi}{(2\pi\hbar)^n} \mathfrak{B}(\xi) Tr[\mathfrak{B}(\xi)\mathfrak{f}] = \mathfrak{f},$$
$$Tr[\mathfrak{B}(\xi)] = 1, \tag{4}$$
$$Tr[\mathfrak{B}(\xi)\mathfrak{B}(\xi')] = (2\pi\hbar)^n \delta^{2n}(\xi - \xi'),$$
$$Tr[\mathfrak{B}(\xi)\mathfrak{B}(\xi')\mathfrak{B}(\xi'')] = (2\pi\hbar)^n \delta^{2n}(\xi'' - \xi) \exp(\frac{i\hbar}{2}\mathcal{P}_{\xi\xi'})(2\pi\hbar)^n \delta^{2n}(\xi - \xi'),$$

$$\mathfrak{B}(\xi)\exp(-\frac{i\hbar}{2}\mathcal{P}_{\xi\xi'})\mathfrak{B}(\xi') = (2\pi\hbar)^n \delta^{2n}(\xi-\xi')\mathfrak{B}(\xi'),$$

where $\mathcal{P}_{\xi\xi'}$ is the Poisson operator

$$\mathcal{P}_{\xi\xi'} = -I^{kl}\frac{\overleftarrow{\partial}}{\partial\xi^k}\frac{\overrightarrow{\partial}}{\partial\xi'^l}.$$

In Equation (4), the first line is obvious. The equations in lines two to five are equivalent to Equations (4.15) – (4.18) in Groenewold (1946). The last equation can be found, e.g., in Krivoruchenko et al. (2006a). The equation in the 6th line is a consequence of the 5th and 7th equations.

The Wigner-Weyl association rule $f(\xi) \leftrightarrow \mathfrak{f}$ takes, in the basis $\mathfrak{B}(\xi)$, the simple form (Groenewold, 1946; Stratonovich, 1957):

$$f(\xi) = Tr[\mathfrak{B}(\xi)\mathfrak{f}], \tag{5}$$

$$\mathfrak{f} = \int \frac{d^{2n}\xi}{(2\pi\hbar)^n} f(\xi)\mathfrak{B}(\xi). \tag{6}$$

In particular, $\mathfrak{r}^i \to \xi^i = Tr[\mathfrak{B}(\xi)\mathfrak{r}^i]$. The reciprocal relation $\xi^i \to \mathfrak{r}^i$, defined by the second line, also holds.

The function $f(\xi)$ can be interpreted as a coordinate of the operator $\mathfrak{f}$ in the basis $\mathfrak{B}(\xi)$, and $Tr[\mathfrak{B}(\xi)\mathfrak{f}]$ can be viewed as the scalar product of $\mathfrak{B}(\xi)$ and $\mathfrak{f}$. Other operator bases are also discussed (Balazs & Jennings, 1984).

The set $Op(L^2(\mathbb{R}^n))$ is closed under the addition and the multiplication of the operators. Thus, vector space $\mathbb{V}$ acquires the structure of associative algebra. For any two functions $f(\xi)$ and $g(\xi)$, a third function can be constructed (Groenewold, 1946):

$$f(\xi) \star g(\xi) \equiv Tr[\mathfrak{B}(\xi)\mathfrak{f}\mathfrak{g}]. \tag{7}$$

The operation is called star-product (★-product) of $f(\xi)$ and $g(\xi)$. The explicit form of ★-product is as follows

$$f(\xi) \star g(\xi) = f(\xi)\exp(\frac{i\hbar}{2}\mathcal{P})g(\xi), \tag{8}$$

where $\mathcal{P} = \mathcal{P}_{\xi\xi}$. The ★-product splits into symmetric and skew-symmetric parts

$$f \star g = f \circ g + \frac{i\hbar}{2}f \wedge g. \tag{9}$$

The skew-symmetric component is known as the Moyal bracket (Groenewold, 1946; Moyal, 1949). In the classical limit, the Moyal bracket $f \wedge g$ turns into the Poisson bracket $\{f,g\} \equiv f(\xi)\mathcal{P}g(\xi)$.

Weyl's symbol of the symmetrised product of the operators of the canonical coordinates and momenta $\mathfrak{r}^{(i_1}\mathfrak{r}^{i_2}...\mathfrak{r}^{i_s)}$ coincides with the dot (ordinary) product of the associated canonical variables

$$Tr[\mathfrak{B}(\xi)\mathfrak{r}^{(i_1}\mathfrak{r}^{i_2}...\mathfrak{r}^{i_s)}] = \xi^{i_1}\xi^{i_2}...\xi^{i_s},$$

which is explicitly symmetric for permutations of the indices. The symmetrised products of the Hermitian operators $\mathfrak{u}^i$ correspond to the symmetrised ★-products of the associated real functions $u^i(\xi) = Tr[\mathfrak{B}(\xi)\mathfrak{u}^i]$:

$$Tr[\mathfrak{B}(\xi)\mathfrak{u}^{(i_1}\mathfrak{u}^{i_2}...\mathfrak{u}^{i_s)}] = u^{(i_1}(\xi) \circ u^{i_2}(\xi) \circ ... \circ u^{i_s)}(\xi).$$

The ∘-product is not associative. Its order here is not important, because the indices are symmetrised.

Weil's map from functions to operators was originally formulated in terms of the Taylor expansion. Consider the decomposition of a function near zero

$$f(\xi) = \sum_{s=0}^{\infty} \frac{1}{s!} \frac{\partial^s f(0)}{\partial\xi^{i_1}...\partial\xi^{i_s}}\xi^{i_1}...\xi^{i_s}.$$

According to the Weyl rule (Weyl, 1927, 1929, 1931), the function $f(\xi)$ maps into the operator $\mathfrak{f}_T$

$$\mathfrak{f}_T \equiv f(\mathfrak{r}) = \sum_{s=0}^{\infty} \frac{1}{s!} \frac{\partial^s f(0)}{\partial\xi^{i_1}...\partial\xi^{i_s}}\mathfrak{r}^{i_1}\mathfrak{r}^{i_2}...\mathfrak{r}^{i_s}.$$

Note that the indices of summation over the coordinate components are automatically symmetrised. A simple calculation,

$$f_T(\xi) = Tr[\mathfrak{B}(\xi)\mathfrak{f}_T] = \sum_{s=0}^{\infty} \frac{1}{s!} \frac{\partial^s f(0)}{\partial\xi^{i_1}...\partial\xi^{i_s}}\xi^{i_1} \star \xi^{i_2} \star ... \star \xi^{i_s}$$

$$= \sum_{s=0}^{\infty} \frac{1}{s!} \frac{\partial^s f(0)}{\partial\xi^{i_1}...\partial\xi^{i_s}}\xi^{i_1}\xi^{i_2}...\xi^{i_s},$$

shows that the Taylor expansion of the product of the operators of the canonical coordinates and momenta provides the correspondence rule, completely equivalent to Equation (6). Thus, we obtain $f_T(\xi) = f(\xi) \leftrightarrow \mathfrak{f}_T = \mathfrak{f}$.

For any operator $\mathfrak{f} \in Op(L^2(\mathbb{R}^n))$, one may find a function $f(\xi)$ such that $\mathfrak{f} = f(\mathfrak{x})$. This property shows the completeness of the set of operators of canonical coordinates and momenta in $Op(L^2(\mathbb{R}^n))$.

The average value of a physical observable $\mathfrak{f}$ is determined by the trace of $\mathfrak{f}\mathfrak{r}$, where $\mathfrak{r}$ is the density matrix, or by averaging the function $f(\xi)$ over the Wigner function

$$W(\xi) = Tr[\mathfrak{B}(\xi)\mathfrak{r}]. \tag{10}$$

Because $Tr[\mathfrak{r}] = 1$, the Wigner function is normalised to unity

$$\int \frac{d^{2n}\xi}{(2\pi\hbar)^n} W(\xi) = 1. \tag{11}$$

The average value of $\mathfrak{f}$ is given by

$$\langle \mathfrak{f} \rangle = Tr[\mathfrak{f}\mathfrak{r}] = \int \frac{d^{2n}\xi}{(2\pi\hbar)^n} f(\xi) \star W(\xi) = \int \frac{d^{2n}\xi}{(2\pi\hbar)^n} f(\xi)W(\xi). \tag{11}$$

In this case, the $\star$-product of $f(\xi)$ and $W(\xi)$ can be replaced by a dot product, because the derivatives of the Poisson operator are reduced to surface integrals and can be omitted.

Not every normalised function in phase space can be interpreted as the Wigner function. The eigenvalues of the density matrix are nonnegative. For each density matrix, one may then find a Hermitian matrix $\mathfrak{r}^{1/2}$, such that $\mathfrak{r}^{1/2}\mathfrak{r}^{1/2} = \mathfrak{r}$. For any Wigner function $W(\xi)$ a function $W_{1/2}(\xi)$ exists, such that $W_{1/2}(\xi) \star W_{1/2}(\xi) = W(\xi)$.

For a pure state, $\mathfrak{r} = |\psi\rangle\langle\psi|$, and the Wigner function equals $W(\xi) = \langle\psi|\mathfrak{B}(\xi)|\psi\rangle$. To show that the Wigner map, which is described in standard textbooks, is equivalent to Equation (5), we first find the mixed matrix elements of the basis vectors of $V$:

$$\langle x|\mathfrak{B}(q,p)|k\rangle = 2^n \exp\left(-\frac{2i}{\hbar}(p-k)\cdot(q-x)\right)\langle x|k\rangle,$$

where $x$ and $q$ are the canonical coordinates and $k$ and $p$ are the canonical momenta. After simple transformations, the usual expression then follows:

$$W(q,p) = \langle\psi|\mathfrak{B}(q,p)|\psi\rangle = \int \frac{d^n x d^n k}{(2\pi\hbar)^n}\langle\psi|x\rangle\langle x|\mathfrak{B}(q,p)|k\rangle\langle k|\psi\rangle$$

$$= \int d^n x \langle\psi|q+\frac{x}{2}\rangle\exp(\frac{i}{\hbar}xp)\langle q-\frac{x}{2}|\psi\rangle.$$

Applying the Schwartz inequality to the integral and taking the normalisation condition $\langle\psi|\psi\rangle = 1$ into account, one obtains the constraint $-2^n \leq W(\xi) \leq 2^n$ (Baker, 1958). The value of $W(\xi)$ is bounded, provided $|\psi\rangle$ has a finite norm, which is the case for bound states of discrete spectrum and wave packets in the continuum.

### 3. Quantum trajectories in phase space as characteristics

One-parameter unitary transformations acting on the operators of the canonical coordinates and momenta generate the trajectories in the phase space by Wigner's association rule. Knowledge of these trajectories is equivalent to knowledge of the quantum dynamics. The time-dependent symbols of the operators are functions of the trajectories. In this sense, the phase-space trajectories play a special role, similar to the role of classical trajectories in solving the Liouville equation.

The Liouville equation is a partial differential equation (PDE). Its general solution can be represented by its characteristics. The characteristics of the classical Liouville equation are the classical trajectories of the particles. Quantum trajectories solve the Groenewold evolution equation. For this reason, we call them "quantum characteristics."

### 3.1 Wigner map of unitary transformation

Consider a unitary transformation acting on the operators $\mathfrak{f} \to \mathfrak{f}' = \mathfrak{U}^+\mathfrak{f}\mathfrak{U}$, where $\mathfrak{U}^+\mathfrak{U} = \mathfrak{U}\mathfrak{U}^+ = 1$. The operators of the canonical coordinates and momenta are transformed according to the rule $\mathfrak{x}^i \to \mathfrak{x}'^i = \mathfrak{U}^+\mathfrak{x}^i\mathfrak{U}$, while their Weyl's symbols are transformed according to the rule

$$\xi^i \to \xi'^i = u^i(\xi) \equiv Tr[\mathfrak{B}(\xi)\mathfrak{U}^+\mathfrak{x}^i\mathfrak{U}]. \tag{12}$$

Thus, unitary transformations in the space $Op(L^2(\mathbb{R}^n))$ generate, through the Wigner association rule, a coordinate transformation in the phase space $\mathbb{R}^{2n}$. Such transformations are not canonical (see below), and we call them "unitary transformations". The transformation law of functions under the unitary transformation takes the form

$$f(\xi) \to f'(\xi) = Tr[\mathfrak{B}(\xi)\mathfrak{f}'] = Tr[\mathfrak{B}(\xi)\mathfrak{U}^+\mathfrak{f}\mathfrak{U}]$$

$$= \sum_{s=0}^{\infty} \frac{1}{s!} \frac{\partial^s f(0)}{\partial \xi^{i_1} \partial \xi^{i_2} ... \partial \xi^{i_s}} Tr[\mathfrak{B}(\xi)\mathfrak{U}^+ \mathfrak{x}^{i_1} \mathfrak{x}^{i_2}...\mathfrak{x}^{i_s}\mathfrak{U}]$$

$$= \sum_{s=0}^{\infty} \frac{1}{s!} \frac{\partial^s f(0)}{\partial \xi^{i_1} \partial \xi^{i_2} ... \partial \xi^{i_s}} Tr[\mathfrak{B}(\xi)\mathfrak{x}'^{i_1} \mathfrak{x}'^{i_2}...\mathfrak{x}'^{i_s}] \quad (13)$$

$$= \sum_{s=0}^{\infty} \frac{1}{s!} \frac{\partial^s f(0)}{\partial \xi^{i_1} \partial \xi^{i_2} ... \partial \xi^{i_s}} u^{i_1}(\xi) \star u^{i_2}(\xi) \star ... \star u^{i_s}(\xi)$$

$$\equiv f(\star u(\xi)).$$

This expression defines a composite ★-function. The ★-product here may be substituted by the ∘-product. The ∘-product contains even powers of the Planck's constant in its decomposition. Consequently, the expansion around $f(u(\xi))$ contains even powers of $\hbar$. Provided that $u(\xi)$ is a linear function, $f(\star u(\xi)) = f(u(\xi))$. In the general case, the composition law of two functions is not local: $f(\star u(\xi)) \equiv f(\circ u(\xi)) \neq f(u(\xi))$.

### 3.2 Conservation of Moyal bracket

The antisymmetrised products of an even number of operators of canonical coordinates and momenta are *c*-numbers. These products are invariant under unitary transformations:

$$\mathfrak{U}^+ \mathfrak{x}^{[i_1} \mathfrak{x}^{i_2}...\mathfrak{x}^{i_{2s}]} \mathfrak{U} = \mathfrak{x}^{[i_1} \mathfrak{x}^{i_2}...\mathfrak{x}^{i_{2s}]}. \quad (14)$$

In the phase space, this equation is expressed as:

$$u^{[i_1}(\xi) \star u^{i_2}(\xi) \star ... \star u^{i_{2s}]}(\xi) = \xi^{[i_1} \star \xi^{i_2} \star ... \star \xi^{i_{2s}]}$$

$$= \left(\frac{-i\hbar}{2}\right)^s \frac{1}{(2s)!} \sum_\sigma (-)^\sigma I^{i_1 i_2}...I^{i_{2s-1} i_{2s}}.$$

The summation is over all of the permutations of the indices. The sign is plus or minus, depending on whether the sequence $\sigma$ is an even or odd permutation of (1,2,...,2*n*). The invariance of these antisymmetrised products of even numbers of operators of canonical coordinates and momenta constitutes a quantum analogue of the Poincaré theorem on the conservation of 2*n* forms in classical Hamiltonian dynamics (Krivoruchenko et al, 2006c). In particular,

$$u^i(\xi) \wedge u^j(\xi) = \xi^i \wedge \xi^j = -I^{ij}. \quad (15)$$

The real functions $u^i(\xi)$ are associated, by virtue of Equation (6), with the Hermitian operators $\mathfrak{x}'^u$. If $u^i(\xi)$ satisfies Equation (15), then the operators $\mathfrak{x}'^u$ obey the commutation relations

$$[u^i(\mathfrak{x}), u^j(\mathfrak{x})] = [\mathfrak{x}'^i, \mathfrak{x}'^j] = -I^{ij}.$$

We are mainly interested in the case in which $\mathfrak{U}$ is the evolution operator. Applying a unitary transformation to the product $\mathfrak{f}\mathfrak{g}$, we obtain the function $f(\zeta) \star g(\zeta)|_{\zeta = \star u(\xi, \tau)}$ that is associated with the expression $\mathfrak{U}^+ \mathfrak{f}\mathfrak{g}\mathfrak{U}$ and the function $f(\star u(\xi)) \star g(\star u(\xi))$ associated with the expression $(\mathfrak{U}^+ \mathfrak{f}\mathfrak{U})(\mathfrak{U}^+ \mathfrak{g}\mathfrak{U})$. These operators coincide, so their symbols coincide:

$$f(\zeta) \star g(\zeta)|_{\zeta = \star u(\xi, \tau)} = f(\star u(\xi)) \star g(\star u(\xi)). \quad (16)$$

In the first case, the ★-product is calculated with respect to $\zeta^i$, and it is calculated with respect to $\xi^i$ in the second case. Equation (16) shows that the ★-product can be calculated in the original coordinate system prior to the change of the variables, or we can first change the variables and then compute the ★-product. Equation (16) holds separately for the symmetric and antisymmetric parts of the ★-product.

Thus, we can calculate the ★-product for any of the unitary equivalent coordinate systems. Dynamic equations constructed using the summation and multiplication of ★-functions are generally covariant under unitary transformations in phase space. The ★-product is not invariant under canonical transformations.

Example: The classical Liouville equation is covariant under canonical transformations. This equation, however, is not covariant under unitary transformations. The quantum Liouville equation, i.e., the Wigner map of the von Neumann equation, is covariant under unitary transformations and is not covariant under canonical transformations.

### 3.3 Phase flow generated by an evolution operator

The one-parameter family of the unitary transformations describe the evolution of quantum systems, and it is usually parameterised in the form

$$\mathfrak{U}(\tau) = \exp(-\frac{i}{\hbar}\mathfrak{H}\tau),$$

where $\mathfrak{H}$ is the Hamiltonian operator. The functions $u^i(\xi)$, defined in Equation (12), acquire the dependence on the parameter $\tau$, so that we can write $u^i(\xi, \tau)$. These functions determine the quantum phase flow, which is a quantum-mechanical analogue of the phase flow in the Hamiltonian formalism of classical mechanics.

Equation (13) shows that the evolution of the symbols of the operators in the Heisenberg representation is completely determined by the functions $u^i(\xi, \tau)$.

We use the term "canonical transformation" in the conventional sense to refer to the coordinate transformations that preserve the Poisson bracket. The transformations preserving the Moyal bracket are unitary transformations. These transformations correspond to the action of a unitary operator in the Hilbert space. The unitary transformation in the phase space represents the quantum deformation of the canonical transformation.

Quantum characteristics arise in Heisenberg's matrix mechanics. Suppose that we have solved the evolution equations for the operators of the canonical coordinates and momenta in the Heisenberg representation. These operators evolve according to $\mathfrak{x}^i \to \mathfrak{x}^i(\tau) = \mathfrak{U}^+(\tau)\mathfrak{x}^i\mathfrak{U}(\tau)$. We use the earlier assertion that, for any operator $\mathfrak{f}$, one can find a function $f(\xi)$ through which $\mathfrak{f}$ is represented in the form $f(\mathfrak{x})$. The same operator $\mathfrak{f}$ at time $\tau$ is equal to

$$\mathfrak{f}(\tau) \equiv \mathfrak{U}^+(\tau)\mathfrak{f}\mathfrak{U}(\tau) = \mathfrak{U}^+(\tau)f(\mathfrak{x})\mathfrak{U}(\tau) = f(\mathfrak{U}^+(\tau)\mathfrak{x}\mathfrak{U}(\tau)) = f(\mathfrak{x}(\tau)).$$

This equation shows that the operators of the canonical coordinates and momenta are characteristics that determine the evolution for all of the operators in $Op(L^2(\mathbb{R}^n))$. This property is fully transferred to the phase space upon deformation quantisation.

### 3.4 Energy conservation and composition law for trajectories

Energy conservation in the process of evolution means

$$H(\xi) = H(\star u(\xi,\tau)), \qquad (17)$$

where $H(\xi) = Tr[\mathfrak{B}(\xi)\mathfrak{H}]$ is the Hamiltonian function of the quantum system. We see that energy is conserved along quantum characteristics, but not in the geometric sense. The $\star$-product sign in the argument indicates the non-local nature of the conservation law.

The law of composition of the particle trajectories also has a non-local character:

$$u(\xi,\tau_1 + \tau_2) = u(\star u(\xi,\tau_1),\tau_2).$$

Such compositions, but without the $\star$-product, are valid for the trajectories of classical particles. The $\star$-product does not allow considering the motion of particles as movement along a certain trajectory in the geometrical sense.

### 3.5 Quantum Hamilton equations

Quantum trajectories can be found by solving Hamilton's equations, which can be written in one of four equivalent forms:

$$\begin{aligned}\frac{\partial}{\partial \tau}u^i(\xi,\tau) &= \{\zeta^i, H(\zeta)\}|_{\zeta=\star u(\xi,\tau)} \\ &= \zeta^i \wedge H(\zeta)|_{\zeta=\star u(\xi,\tau)} \\ &= u^i(\xi,\tau) \wedge H(\star u^i(\xi,\tau)) \\ &= u^i(\xi,\tau) \wedge H(\xi),\end{aligned} \qquad (18)$$

with the initial conditions

$$u^i(\xi,0) = \xi^i. \qquad (19)$$

These equations appear as Wigner's image of the evolution equations for operators of the canonical coordinates and momenta in the Heisenberg representation. The equivalence of the different records of the right-hand side can be verified with the help of the above-described properties of the $\star$-product, the rules of substitution (Equation (16)), and the condition of energy conservation (Equation (17)). Note that $\{\zeta^i, H(\zeta)\} = \zeta^i \wedge H(\zeta) = H(\zeta)^i$.

The substitution $\zeta = \star u(\xi,\tau)$ in the first line of Equation (18) leads to a modification of the classical expression for the right-hand side and, correspondingly, to quantum deformation of the classical phase flow. The value of $\partial u^i(\xi,\tau) / \partial \tau$ depends on the phase space coordinate $u^i(\xi,\tau)$, as in classical mechanics, and on the infinite number of partial derivatives of $u^i(\xi,\tau)$ as a specific manifestation of the quantum non-locality.

An equivalent form of Equation (18), using a cluster expansion of the $\star$-exponentials, was given by Osborn & Molzahn (1995). Equation (18) was found independently by Krivoruchenko & Faessler (2006b).

### 3.6 Quantum Liouville equation

The functions corresponding to physical observables evolve in the Heisenberg representation according to the equation

$$f(\xi,\tau) = Tr[\mathfrak{B}(\xi)\mathfrak{U}^+(\tau)\mathfrak{f}\mathfrak{U}(\tau)] = f(\star u(\xi,\tau),0), \qquad (19)$$

while the Wigner function remains constant. The evolution law can be expressed in terms of Green's function in the phase space as

$$f(\xi,\tau) = \int \frac{d^{2n}\eta}{(2\pi\hbar)^n} G(\xi,\eta,\tau) f(\eta,0).$$

With the help of quantum characteristics, a compact expression for Green's function can be written as

$$G(\xi,\eta,\tau) = (2\pi\hbar)^n \delta^{2n}(\star u(\xi,\tau) - \eta).$$

The function $f(\xi,\tau)$ obeys the Groenewold evolution equation (Groenewold, 1946)

$$\frac{\partial}{\partial \tau} f(\xi,\tau) = f(\xi,\tau) \wedge H(\xi), \qquad (20)$$

which is the Wigner map of the evolution equation of the operator $\mathfrak{f}$ in the Heisenberg representation. The right-hand side can be replaced by the equivalent expressions $f(\xi,\tau) \wedge H(\star u(\xi,\tau))$ or $f(\zeta,0) \wedge H(\zeta)|_{\zeta=\star u(\xi,\tau)}$.

The solutions to the evolution equations for the quantum characteristics and functions $f(\xi,\tau)$ can be represented as a formal power series in the parameter $\tau$:

$$u^i(\xi,\tau) = \sum_{s=0}^{\infty} \frac{\tau^s}{s!} \underbrace{(...((\xi^i \wedge H(\xi)) \wedge H(\xi)) \wedge ... H(\xi))}_{s},$$

$$f(\star u(\xi,\tau)) = \sum_{s=0}^{\infty} \frac{\tau^s}{s!} \underbrace{(...((f(\xi) \wedge H(\xi)) \wedge H(\xi)) \wedge ... H(\xi))}_{s}. \tag{21}$$

Unitary coordinate transformations are canonical to the first order in $\tau$ (Dirac, 1930; Weyl, 1931).

For higher orders, deviations from the canonicity arise (Krivoruchenko & Faessler, 2006b). Until recently, these deviations were not well understood.[1] The infinitesimal transformations generate canonical or unitary global transformations depending on how we define the multiplication. If this is the usual dot product, then we obtain the canonical transformations. If this is the ★-product, then we obtain unitary transformations.

If an operator $\mathfrak{A}$ commutes with $\mathfrak{H}$, then its symbol is preserved in the sense of $A(\xi) = A(\star u(\xi,\tau))$. The density matrix of a stationary state commutes with $\mathfrak{H}$; therefore, in the Schrödinger representation $W_S(\xi) = W_S(\star u(\xi,-\tau))$, the Wigner function does not evolve.

In the harmonic oscillator, the quantum trajectory depends linearly on $\xi^i$ and coincides with the classical trajectory. In this case, the ★-symbol in the argument of the Wigner function can be omitted, and we can write $W_S(\xi) = W_S(u(\xi,-\tau))$ for a stationary state and $W_S(\xi,\tau) = W_S(u(\xi,-\tau),0)$ for an arbitrary state.

## 4. Semiclassical expansion of quantum characteristics

The methods of this section apply to the evolution problem. The most promising applications appear to be connected to many-body scattering and transport models.

The problem of evolution can by divided into two parts. First, we seek a solution to the quantum Hamilton equations. Second, we use Equation (19) to find the time-dependent symbols of the operators. The key issue is an efficient algorithm to calculate the ★-functions. These functions arise in the quantum Hamilton equations and in solving the evolution problem for functions.

### 4.1 Semiclassical expansion of ★-functions

---

[1] In the papers by B. Leaf, J. Math. Phys. **9**, 769 (1968) and T. Curtright and C. Zachos, J. Phys. A 32, 771 (1999), erroneous statements about the entire coincidence of classical and quantum trajectories can be found.

We consider the semiclassical expansion of $f(\star u(\xi,\tau))$ around $f(u(\xi,\tau))$. The function $f(\xi)$ can be represented through its Fourier transform

$$f(\xi) = \int \frac{d^{2n}\eta}{(2\pi\hbar)^n} \exp(\frac{i}{\hbar}\eta_k \xi^k) f(\eta). \tag{22}$$

Determining how to calculate $\exp(\star U)$, where $U = \frac{i}{\hbar}\eta_k u^k(\xi,\tau)$, is sufficient. With the help of Equation (13), we find

$$\exp(\star U) = \left(1 + \hbar^2 c_2 + \hbar^4 c_4 + O(\hbar^6)\right)\exp(U),$$

where

$$c_2 = -\frac{1}{48}(2UUP^2U + 3UP^2U),$$

$$c_4 = \frac{1}{23040}(90(UP^2U)UP^2U + 60(UP^2U)P^2U + 48(UUP^2U)UP^2U + 45(UP^2U)(UP^2U) \tag{23}$$
$$+ 60(UUP^2U)(UP^2U) + 20(UUP^2U)(UUP^2U) + 30(UUP^2U)P^2U)$$

$$+ \frac{1}{11520}(6UUUUP^4U + 45UUUP^4U + 30(UUP^4U)_1 + 40(UUP^4U)_2 + 15UP^4U).$$

The operator $P$ acts as follows:

$$AP^2B = A_{,kl}B^{,kl}, \quad ABP^2C = A_{,k}B_{,l}C^{,kl}, \quad AP^4B = A_{,ijkl}B^{,ijkl},$$
$$(ABP^4C)_1 = A_{,ij}B_{,kl}C^{,ijkl}, \quad (ABP^4C)_2 = A_{,i}B_{,jkl}C^{,ijkl},$$
$$ABCP^4D = A_{,i}B_{,j}C_{,kl}D^{,ijkl}, \quad ABCDP^4E = A_{,i}B_{,j}C_{,k}D_{,l}E^{,ijkl},$$

where

$$A(\xi)_{,i_1...i_s} = \frac{\partial^s A(\xi)}{\partial \xi^{i_1}...\partial \xi^{i_s}}, \quad A(\xi)^{,i_1...i_s} = A(\xi)_{,j_1...j_s} I^{j_1 i_1}...I^{j_s i_s}. \tag{24}$$

Osborn & Molzahn (1995) and Gracia-Saz (2004) developed a diagram technique to calculate Weyl's symbols of composite operators for higher orders of the $\hbar$-expansion. Equation (23) from Krivoruchenko et al. (2006c), obtained with the use of MAPLE, agrees with the calculation of Gracia-Saz (2004).

The expansion of $f(\star u(\xi,\tau))$ is now straightforward. We replace $\eta_i \to -i\hbar\partial/\partial u^i$ and $U \to u^i(\xi)\partial/\partial u^i$ to obtain

$$f(\star u(\xi,\tau)) = f(u(\xi,\tau)) - \frac{\hbar^2}{24} u^i(\xi,\tau)_{,l} u^j(\xi,\tau)_{,m} u^k(\xi,\tau)^{,lm} f(u(\xi,\tau))_{,ijk} \qquad (25)$$
$$- \frac{\hbar^2}{16} u^i(\xi,\tau)_{,kl} u^j(\xi,\tau)^{,kl} f(u(\xi,\tau))_{,ij} + O(\hbar^4).$$

The derivatives of $f(u(\xi,\tau))$ are calculated with respect to $u$.

As a simple application, one can find a semiclassical expansion of the Weyl symbol for the finite-temperature density matrix (Wigner, 1932):

$$W(\xi) \propto \exp(-\frac{\star H(\xi)}{T})$$
$$= \exp(-\frac{H(\xi)}{T}) \left( 1 + \frac{\hbar^2}{24T^3} H(\xi)_{,l} H(\xi)_{,m} H(\xi)^{,lm} - \frac{\hbar^2}{16T^2} H(\xi)_{,kl} H(\xi)^{,kl} + O(\hbar^4) \right).$$

After the replacement $1/T \to i\tau/\hbar$, the expression for the Weyl symbol of the evolution operator $\mathfrak{U}(\tau) = \exp(-\frac{i}{\hbar}\mathfrak{H}\tau)$ is derived.

The evolution operator and its Weyl's symbol are singular as $\hbar \to 0$. The expansion of the semiclassically admissible ★-functions starts, however, with a classical expression, which is independent of $\hbar$. The question of what happened to the singularity arises. The answer is obtained by considering the time-dependent operator $\mathfrak{f}(\tau) = \mathfrak{U}^+(\tau)\mathfrak{f}\mathfrak{U}(\tau)$. The derivatives of $\mathfrak{f}(\tau)$ of order $k$ are expressed in terms of $k$ commutators. Each commutator generates $\hbar$. The commutator $[\frac{i}{\hbar}\mathfrak{H},\mathfrak{f}]$ and the high-order terms $[\frac{i}{\hbar}\mathfrak{H},[\frac{i}{\hbar}\mathfrak{H},...,[\frac{i}{\hbar}\mathfrak{H},\mathfrak{f}]...]]$ are regular as $\hbar \to 0$. Therefore the Weyl symbol of $\mathfrak{f}(\tau)$ has a classical limit.

In transport models, the solutions to the evolution problem are based on solving systems of ODEs. We now turn our attention to the construction of the ODEs.

### 4.2 Semiclassical expansion of quantum Hamilton equations

Suppose we have a system of $N$ particles. The interaction of the particles is described by some potential. The initial-state wave function is assumed to be known. Consequently, the initial-state Wigner function is known.

The first step in solving the evolution problem consists of finding the quantum characteristics, using Equation (18). We expand the solution in powers of Planck's constant

$$u^i(\xi,\tau) = \sum_{r=0}^{\infty} \hbar^{2r} u_r^i(\xi,\tau). \qquad (25)$$

Here, $u_0^i(\xi,\tau)$ is the classical trajectory starting at time $\tau = 0$ at $\xi^i \in \mathbb{R}^{2n}$. The initial conditions for the quantum corrections $r \geq 1$ at $\tau = 0$ are then set equal to zero. As a result,

$$u_0^i(\xi,0) = \xi^i, \quad r = 0, \qquad (26)$$
$$u_r^i(\xi,0) = 0, \quad r \geq 1.$$

The right side of Equation (18) is the ★-function, so we use the decomposition

$$F^i(\star u(\xi,\tau)) \equiv \{\zeta^i, H(\zeta)\}|_{\zeta=\star u(\xi,\tau)} = \sum_{r=0}^{\infty} \hbar^{2r} F_r^i(u_0(\xi,\tau),...,u_r(\xi,\tau)). \qquad (27)$$

If the functions $u^i(\xi,\tau)$ are known, then the functions $F_r^i(u_0(\xi,\tau),...,u_r(\xi,\tau))$ are completely determined. $F_r^i(u_0(\xi,\tau),...,u_r(\xi,\tau))$ also depends on the derivatives of $u_0(\xi,\tau),...,u_r(\xi,\tau)$ with respect to $\xi^i$. In particular,

$$F_0^i(u_0) = F^i(u_0),$$
$$F_1^i(u_0, u_1) = u_1^j(\xi,\tau) F^i(u_0)_{,j} - \frac{1}{24} u_0^j(\xi,\tau)_{,m} u_0^k(\xi,\tau)_{,n} u_0^l(\xi,\tau)^{,mn} F^i(u_0)_{,jkl} \qquad (28)$$
$$- \frac{1}{16} u_0^j(\xi,\tau)_{,lm} u_0^k(\xi,\tau)^{,lm} F^i(u_0)_{,jk}.$$

### 4.3 Jacobi fields

The second line of Equation (28) contains the first- and second-order derivatives of $u_0^i(\xi,\tau)$ with respect to $\xi^i$. Therefore, we should monitor the evolution of the trajectories and their derivatives with respect to the initial coordinates

$$J^i_{r,j_1...j_t}(\xi,\tau) = \frac{\partial^t u_r^i(\xi,\tau)}{\partial \xi^{j_1}...\partial \xi^{j_t}}. \qquad (29)$$

These values determine the decomposition of $f(\star u(\xi,\tau))$ and determine the high-order quantum corrections to the phase-space trajectories.

We call these values "Jacobi fields". This term is adopted in Hamiltonian mechanics for the first-order derivatives of $u_0^i(\xi,\tau)$, that determine the stability of the systems. The value of $J^i_{0,j}(\xi,\tau) = \partial u_0^i(\xi,\tau)/\partial \xi^j$ is known as the Jacobi matrix, and its determinant is called the Jacobian. In quantum mechanics, the derivatives of higher orders are involved. In the following, the number $r$ is called the order of the Jacobi field, the number of lower indices $t$ is called the degree of Jacobi field.

In the notation of Equation (29), the first pair of evolution equations becomes

$$\frac{\partial}{\partial \tau} u_0^i(\xi,\tau) = F^i(u_0),$$

$$\frac{\partial}{\partial \tau} u_1^i(\xi,\tau) = u_1^j(\xi,\tau) F^i(u_0)_{,j} - \frac{1}{24} J_{0,m}^j(\xi,\tau) J_{0,n}^k(\xi,\tau) J_0^{l,mn}(\xi,\tau) F^i(u_0)_{,jkl} \qquad (30)$$

$$- \frac{1}{16} J_{0,lm}^j(\xi,\tau) J_0^{k,lm}(\xi,\tau) F^i(u_0)_{,jk}.$$

In the first line, the classical Hamilton equations are recognisable. The second and third lines determine the lowest-order quantum correction to the classical trajectory.

The system in Equation (30) is not yet closed. It needs to be supplemented by the equations of motion of $J_{0,j}^i(\xi,\tau)$ and $J_{0,jk}^i(\xi,\tau)$. These equations are obtained by taking the first- and second-order derivatives of the first equation with respect to the initial coordinates:

$$\frac{\partial}{\partial \tau} J_{0,j}^i(\xi,\tau) = F^i(u_0)_{,k} J_{0,j}^k(\xi,\tau),$$

$$\frac{\partial}{\partial \tau} J_{0,jk}^i(\xi,\tau) = F^i(u_0)_{,lm} J_{0,j}^l(\xi,\tau) J_{0,k}^m(\xi,\tau) + F^i(u_0)_{,l} J_{0,jk}^l(\xi,\tau). \qquad (31)$$

Differentiating Equation (26) on $\xi^i$, we obtain the initial conditions for the Jacobi fields. In general,

$$J_{r,j}^i(\xi,0) = \delta^i_{\ j}, \quad r = 0,$$
$$J_{r,j_1...j_t}^i(\xi,0) = 0, \quad r = 0, t \geq 2 \text{ or } r \geq 1, t \geq 1. \qquad (32)$$

The coordinates $\xi^i$ enter Equations (30) and (31) as parameters. Thus, we meet a typical case of first-order ODEs, for the variables $u_0^i(\xi,\tau)$, $u_1^i(\xi,\tau)$, $J_{0,j}^i(\xi,\tau)$ and $J_{0,jk}^i(\xi,\tau)$ with the initial conditions given in Equations (26) and (32).

Here, we show a proof of the statement (Krivoruchenko et al., 2006c) that, in any fixed order of $\hbar$, we still have a finite system of first-order ODEs for the variables $u_r^i(\xi,\tau)$ and the associated Jacobi fields.

### 4.4 Reduction of the quantum Hamilton equations to a finite system of first-order ODEs

Let us consider the effect of the ★-product. According to Equation (8), each power of $\hbar$ is accompanied by differentiation. For order $\hbar^{2s}$, expansion of $F^i(\star u(\xi,\tau))$ contains the derivatives of order $2s$ at most. Therefore, the Jacobi fields have the highest degree $2s$ ($1 \leq t \leq 2s$).

This assertion can be strengthened. The number of indices $t$ in reality also depends on the order $r$ of the Jacobi fields $J_{r,j_1...j_t}^i(\xi,\tau)$. In the expansion in Equation (13), the derivatives of the trajectory $u_0^i(\xi,\tau)$ of order $\leq 2s$, the derivatives of the trajectory $u_1^i(\xi,\tau)$ of order $\leq 2s - 2$, ..., and the derivatives of the trajectory $u_{s-1}^i(\xi,\tau)$ of order $\leq 2$ survive among all of the derivatives of orders $\leq 2s$. The highest-order correction $u_s^i(\xi,\tau)$ has no derivatives. Thus, the maximum number of lower indices of $J_{r,j_1...j_t}^i(\xi,\tau)$, involved in the expansion to order $\hbar^{2s}$, depends on $r$ and equals $2s - 2r$ ($1 \leq t \leq 2s - 2r$).

Let us consider in more detail the equations of evolution to order $\hbar^{2s}$ for a fixed $r \leq s$.
The first part of the ODE system can be written as

$$\frac{\partial}{\partial \tau} u_r^i = F_r^i(u_0, u_1, ..., u_r, J_0, J_1, ..., J_{r-1}), \qquad (33)$$

where the index $r$ takes the values $1 ... s$. In the argument for the function on the right-hand side, we dropped the indices of the trajectories and the Jacobi fields. Note that the time derivative depends on the Jacobi fields of order $r - 1$ at most.

The higher-order corrections depend on the lower-order corrections. In the right-hand side of Equation (33), the Jacobi fields have the following degrees: $J_{0,j_1...j_t}^i(\xi,\tau)$ - not more than $2r$, $J_{1,j_1...j_t}^i(\xi,\tau)$ - not more than $2r - 2$, and so on. In the highest order $\hbar^{2s}$-term, the maximum degree of $J_{r-1,j_1...j_t}^i(\xi,\tau)$, entering the right-hand side, is equal to $2s - 2r + 2$. The functions $F_r^i(u_0, u_1, ..., u_r, J_0, J_1, ..., J_{r-1})$ do not depend on the variables with $r'$ for $r < r'$. Equation (33) clearly allows the determination of the trajectories $u_0(\xi,\tau), ..., u_s(\xi,\tau)$, provided that the Jacobi fields are known.

We now supplement the resulting system (Equation (33)) with the equations of evolution of the Jacobi fields.

Consider first Equation (33) for $r = 0$, i.e., the classical Hamilton equations. The right-hand side depends only on $u_0^i(\xi,\tau)$. Differentiating this equation from one to $2s$ times, we obtain evolution equations for the zero-order Jacobi fields of degrees $t = 1 ... 2s$. Equation (33) for $r = 0$ and the $2s$ of these equations form a closed system of ODEs whose solutions are well defined.

As a next step we consider Equations (33) for $r = 1$. The right-hand side depends on the trajectories $u_0^i(\xi,\tau)$ and $u_1^i(\xi,\tau)$ and the Jacobi fields $J_{0,j_1...j_t}^i(\xi,\tau)$ for $t = 1, 2$. Differentiating this equation with respect to $\xi^i$ from one to $2s - 2$ times, we obtain the evolution equations for the first-order Jacobi fields $J_{1,j_1...j_t}^i(\xi,\tau)$ of degrees $t = 1 ... 2s - 2$. After differentiating, the right-hand side depends on the Jacobi fields $J_{0,j_1...j_t}^i(\xi,\tau)$ of degrees $\leq 2s$ (= 2 + 2s - 2), while the Jacobi fields $J_{1,j_1...j_t}^i(\xi,\tau)$ arising from differentiating $u_1^i(\xi,\tau)$ do not have degrees that are higher than the number of derivatives taken, i.e., not above $2s - 2$. Thus, to determine $J_{1,j_1...j_t}^i(\xi,\tau)$, additional information about the zero-order Jacobi fields is not required. From these equations, we can find $u_1^i(\xi,\tau)$ and $J_{1,j_1...j_t}^i(\xi,\tau)$ of degrees $t \leq 2s - 2$.

Further arguments are fairly obvious. We are moving in the direction of increasing the order of Jacobi fields. Consider the general case. We take the derivatives of Equation (33) from one to $2s - 2r$ times and obtain the evolution equations for the order-$r$ Jacobi fields of degrees $t = 1 ... 2s - 2r$.

$$\frac{\partial}{\partial \tau} J^i_{r,j_1\cdots j_t}(\xi,\tau) = G^i_r(u_0, u_1, \ldots, u_r, J_0, J_1, \ldots, J_r). \quad (34)$$

Consider the right-hand side of Equation (33). It depends on the Jacobi fields $J^i_{0,j_1\cdots j_t}(\xi,\tau)$ of degrees $\leq 2r$. After differentiating one to $2s - 2r$ times, a dependence on the Jacobi fields $J^i_{0,j_1\cdots j_t}(\xi,\tau)$ of degrees $\leq 2s$ (= $2r + 2s - 2r$) is acquired. Consequently, for any $r$, the right-hand side of Equations (34) depends on the zero-order Jacobi fields of degree $2s$ at most. Furthermore, the right-hand side of Equation (33) depends on the first-order Jacobi fields of degrees $\leq 2r - 2$. After the differentiation, a dependence on the first-order Jacobi fields of degrees $\leq 2s - 2$ (= $2r - 2 + 2s - 2r$) occurs in Equation (34). The upper value is also independent of $r$.

For a fixed $r$, Equation (33) depends on the $h$–order Jacobi fields ($h < r$) of degrees $t \leq 2r - 2h$. Thus, the evolution equations for the Jacobi fields $J^i_{r,j_1\cdots j_t}(\xi,\tau)$ of degrees $t = 1 \ldots 2s - 2r$ contain the trajectory functions $u_0(\xi,\tau),\ldots,u_r(\xi,\tau)$ and the Jacobi fields $J^i_{h,j_1\cdots j_t}(\xi,\tau)$ of orders $h = 0 \ldots r$ and degrees $t = 1 \ldots 2s - 2h$.

The truncation of the expansion at any $s$ provides us with the complete system of first-order ODEs for the trajectories and the Jacobi fields. The system is determined by Equations (33) and (34) with the initial conditions given in Equations (26) and (32).

In terms of the mathematical induction, the above arguments indicate that Equations (33) and (34) are sufficient to determine the trajectories and the Jacobi fields for some $r$, provided that the trajectories and the Jacobi fields of lower orders $< r$ are determined. We have seen that this is true for $r = 0, 1$; therefore, it is true for all $r$.

**4.5 Perspectives of transport models**

We show the lists of the dynamical variables and the numbers of the independent degrees of freedom in the decomposition of the quantum Hamilton equations up to the fourth order in $\hbar$:

$$\hbar^0: \quad u^i_0(\xi,\tau)$$
$$\to 2n$$
$$\hbar^2: \quad u^i_0(\xi,\tau) \& J^i_{0,j}(\xi,\tau), J^i_{0,jk}(\xi,\tau)$$
$$u^i_1(\xi,\tau)$$
$$\to 2n(2 + 3n + 2n^2)$$
$$\hbar^4: \quad u^i_0(\xi,\tau) \& J^i_{0,j}(\xi,\tau), J^i_{0,jk}(\xi,\tau), J^i_{0,jkl}(\xi,\tau), J^i_{0,jklm}(\xi,\tau)$$
$$u^i_1(\xi,\tau) \& J^i_{1,j}(\xi,\tau), J^i_{1,jk}(\xi,\tau)$$
$$u^i_2(\xi,\tau)$$
$$\to n(18 + 43n + 47n^2 + 20n^3 + 4n^4)/3$$

Here, $n = 3N$, where $N$ is the number of particles. In higher orders, lists of each line are extended by two units at the expense of the Jacobi fields of higher degrees. A new line containing the next order correction to quantum trajectory is also added.

Table 1 shows the number of dynamic degrees of freedom for orders $\hbar^0$, $\hbar^2$, and $\hbar^4$ for the potential scattering of a proton (a nucleus) with nuclei. The spin degrees of freedom of the nucleons are disregarded.

| Reaction | $\hbar^0$ | $\hbar^2$ | $\hbar^4$ |
|---|---|---|---|
| $^2$H + $^2$H | 24 | 7 824 | 499 224 |
| $^9$Be + $^9$Be | 108 | 647 568 | 671 416 128 |
| p + $^{238}$U | 1 434 | 1 477 494 654 | 254 426 548 725 264 |
| $^{238}$U + $^{238}$U | 2 856 | 11 660 059 824 | 7 945 116 177 770 184 |

Table 1. The number of dynamical degrees of freedom for the potential scattering of protons (nuclei) with nuclei of orders $\hbar^0$, $\hbar^2$, and $\hbar^4$.

As the number of Jacobi fields involved in the dynamics increases rapidly with the increasing order of the expansion, limitations due to the semiclassical expansion being restricted by computing power must be considered. Since the mid 1960's, the performance of computers has doubled approximately every 15 months. Due to technical peculiarities of processor manufacturing, this regime is expected to continue for 5 - 15 years.

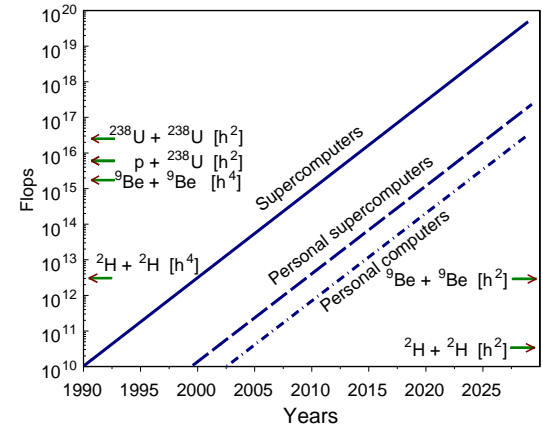

Figure 1. Computer power in flops (floating points operations) required to simulate the proton-nucleus and nucleus-nucleus collisions listed in Table 1 as compared with the growth in power of supercomputers (solid line), personal supercomputers (dashed line) and personal computers (dot-dashed line) starting from 1990. The orders $\hbar^2$ and $\hbar^4$ of the simulations are shown in square brackets.

The theoretical foundations of the transport models used to simulate heavy-ion collisions were created in the late 1980s - early 1990s (Sorge et al., 1989; Aichelin, 1991; Faessler, 1992; Blaettel et al., 1993). Many uncertainties exist in the estimates of the times of the simulations. We assume that simulations of $^{238}$U + $^{238}$U collisions with a supercomputer required about one week in 1990. The calculations are restricted by zero order in the Planck's constant; hence, Jacobi fields are not involved. We are dealing with 2856 dynamical degrees of freedom, as indicated in Table 1, plus the spin degrees of freedom of the nucleons and, potentially, the number of meson degrees of freedom which depends on the beam energy. The calculation time appears to grow linearly with the number of degrees of freedom. Hence, we obtain the estimates shown in Figure 1.

$^2$H + $^2$H reactions of order $\hbar^4$ and $^9$Be + $^9$Be reactions of order $\hbar^2$ could be simulated with supercomputers in 1999. $^9$Be + $^9$Be reactions of order $\hbar^4$, p + $^{238}$U and $^{238}$U + $^{238}$U reactions of order $\hbar^2$ should be simulated with supercomputers by 2011 - 2016. A delay will occur, if computing power is limited to the use of personal supercomputers and computers.

Since the early 1990s, computing power has increased by about *five* orders of magnitude. This dramatic rise in computing power makes it possible to include Jacobi fields in the collision dynamics, to extend beyond the purely classical treatment of phase-space trajectories, currently adopted in all of the transport models.

## 5. Averaging over the Wigner function using the Monte Carlo method

The reduction of the evolution problem to the search for quantum trajectories and the associated Jacobi fields makes it possible to calculate averages using the Monte Carlo method. The problem becomes a task of statistical physics with the modified rules of calculations of probabilities and average values.

The average of the observable associated with an operator $\mathfrak{f}$ at time $\tau$ can be found from the evolution equation (Equation (19)) for the associated function $f(\xi)$. We then use the decomposition (Equation (13)). In the Heisenberg representation, the average is determined by the integral of $f(\xi)$ multiplied by the Wigner function, given at the initial time $W(\xi,0) \equiv W(\xi)$

$$\langle f(\xi,\tau) \rangle = \int \frac{d^{2n}\xi}{(2\pi\hbar)^n} f(\star u(\xi,\tau))W(\xi). \tag{35}$$

We partition the phase space $\mathbb{R}^{2n}$ into two regions $\Omega_+$ and $\Omega_-$, so that $\Omega_+ \cup \Omega_- = \mathbb{R}^{2n}$, in which the Wigner function is positive and negative, respectively. Therefore, we have $W(\xi) = W_+(\xi) - W_-(\xi)$ with $W_\pm(\xi) \geq 0$ in $\Omega_\pm$. Outside these regions, the functions $W_\pm(\xi)$ vanish. As a next step, we generate events in $\Omega_\pm$ (i.e., we select the points $\xi^i \in \Omega_\pm \subset \mathbb{R}^{2n}$), distributed according to the normalised probability densities $W_\pm(\xi)/W_\pm$, with

$$W_\pm = \int \frac{d^{2n}\xi}{(2\pi\hbar)^n} W_\pm(\xi). \tag{35}$$

First consider the region $\Omega_+$. We generate $2n + 1$ numbers $(\xi^i, \gamma)$ with the values of $\xi^i$ uniformly distributed in $\Omega_+$ and the value of $\gamma$ uniformly distributed in the interval $(0, W_{+\max}/W_+)$, where $W_{+\max} = \max_\xi(W_+(\xi))$. If the joint probability density of the variables $(\xi^i, \gamma)$ is given by $\theta(W_+(\xi)/W_+ - \gamma)$, then the marginal probability density of $\xi^i$ equals

$$W_+(\xi)/W_+ = \int_0^{W_{+\max}/W_+} \theta(W_+(\xi)/W_+ - \gamma) d\gamma. \tag{35}$$

To obtain a sample $\{\xi_a^i\}$ of events $a = 1,...,N_+$ distributed with the probability density $W_+(\xi)/W_+$, it is straightforward to discard those generated numbers that do not satisfy the condition $W_+(\xi)/W_+ > \gamma$.

Suppose we generated the numbers $(\xi^i, \gamma)$ at some step. If $W_+(\xi)/W_+ > \gamma$ holds, then we shift the number $N_+$ of successful tests by one unit and assign $\xi_{+a}^i = \xi^i$ for $a = N_+$. Next, we calculate the quantum trajectory $u^i(\xi_{+a}, \tau)$ and the associated Jacobi fields, find the value of $f(\star u(\xi_{+a}, \tau))$ in the required order of the $\hbar$-expansion and store the information. If the inequality $W_+(\xi)/W_+ > \gamma$ is not satisfied, then the event is simply discarded, and we generate the next set of numbers $(\xi^i, \gamma)$. The saved values $\{\xi_a^i\}$ are distributed with the probability density $W_+(\xi)/W_+$. A similar procedure applies to the $\Omega_-$ region.

Suppose we have generated $N_+$ and $N_-$ successful events $\xi_{\pm a} \in \Omega_\pm$. To find the average value of $f(\xi,\tau)$, the values $f(\star u(\xi_{\pm a}, \tau))$ should be multiplied by $W_\pm$, divided by the number of successful tests and summed to give the following:

$$\langle f(\xi,\tau) \rangle \approx \frac{W_+}{N_+} \sum_{a=1}^{N_+} f(\star u(\xi_{+a},\tau)) - \frac{W_-}{N_-} \sum_{a=1}^{N_-} f(\star u(\xi_{-a},\tau)). \tag{36}$$

This equation completes the reduction of the quantum evolution problem to the problem of calculating the statistical averages over an ensemble of quantum characteristics and associated Jacobi fields.

## 6. The scattering problem

In the scattering problem, elementary particles and bound states are considered on an equal footing. We fix the in- and out- scattering states at $\tau = \pm\infty$ and define clusters $\alpha$ of elementary particles and bound states with momenta $p_\alpha'$ and $p_\alpha''$ in the initial and final asymptotic states, respectively. Wigner functions have the form

$$W_{in}(\xi) = \prod_a (2\pi\hbar)^3 \delta(\boldsymbol{p}_a' - \sum_{i\in a}\boldsymbol{p}_i)W_a'(\xi_a),$$
$$W_{out}(\xi) = \prod_a (2\pi\hbar)^3 \delta(\boldsymbol{p}_a'' - \sum_{i\in a}\boldsymbol{p}_i)W_a''(\xi_a). \tag{37}$$

Here, $\xi^i = (q^1,...,q^n, p_1,...,p_n)$, $\xi_\alpha$ are the variables of the particles in cluster $\alpha$. Each cluster contains $\sum_{i\in\alpha} 1$ particles, in total $\sum_\alpha \sum_{i\in\alpha} 1 = N$. A similar situation holds for the partition of particles in the final state.

The Wigner functions $W_{in}(\xi)$ and $W_{out}(\xi)$ are constructed as products of the asymptotic Wigner functions of non-interacting elementary particles and bound states. On the right sides of Equation (37), $W_a'(\xi_\alpha) = W_a''(\xi_\alpha) = 1$ for elementary particles, while Wigner functions of bound states must be constructed on the basis of the wave functions of the bound states of the clusters.

The transition probability is the square modulus of the S-matrix element $w_{fi} = |\langle out | in \rangle|^2$. In terms of the Wigner function,

$$w_{fi} = \int \frac{d^{2n}\xi}{(2\pi\hbar)^n} W_{out}(\star u(\xi,\tau)) W_{in}(\xi), \tag{38}$$

where $n = 3N$. The technique described in the previous sections is fully applicable to Equation (38), and it applies to the scattering problem.

## 7. Conclusion

In this Chapter, we discussed the basic properties of the formalism of deformation quantisation and its applications to the description of the evolution of many-body systems in terms of the expansion in powers of the Planck's constant.

We described the dynamics of quantum systems in phase space. In our presentation, a special role is assigned to quantum trajectories $u^i(\xi,\tau)$, which appear as the Weyl symbols of the operators of the canonical coordinates and momenta in the Heisenberg representation. These trajectories differ from the classical trajectories and from the de Broglie - Bohm trajectories. The transformation of the coordinate system in phase space, associated with quantum trajectories, preserves the Moyal bracket and does not preserve the Poisson bracket. In this sense, the quantum trajectories and phase flow, which they define, can be regarded as a quantum deformation of classical trajectories and phase flow in the formalism of classical Hamiltonian mechanics. The quantum trajectories satisfy the quantum Hamilton equations that are infinite-order PDEs.

Deformation quantisation preserves many features of classical Hamiltonian mechanics. The classical Hamilton's equations are the characteristics equations of the classical Liouville equation for particle distributions in phase space. Accordingly, the solutions of the Hamilton equations contain all of the dynamic information needed to determine the time dependence of all of the observables, including the distribution function.

The same situation occurs in quantum physics. Solutions to the quantum Hamilton equations define quantum trajectories, which possess all of the properties of characteristics. As a rule, characteristics satisfy a system of first-order ODEs, for example, the system of Hamilton's equations. Characteristics are used further to construct solutions of first-order PDEs, such as the classical Liouville equation. The peculiarity of quantum mechanics is that quantum trajectories obey infinite-order PDEs, and they also solve evolution equations that are infinite-order PDEs.

In the Heisenberg representation, the evolution of the Weyl symbol of operator can be written as

$$f(\xi,\tau) = f(\star u(\xi,\tau), 0).$$

The relationships of fundamental interest, such as the one shown above, are formulated in terms of the ★-functions that are not local and form a special class of functionals. To date, no effective methods exist for calculating ★-functions, with the possible exception of the expansion in powers of the Planck's constant. We have outlined recipes to eliminate the ★-symbols from the arguments of composite functions using the semiclassical expansion.

For any fixed order of the semiclassical expansion, the quantum characteristics are constructed by solving a finite system of first-order ODEs. This important circumstance makes it possible to approach the problem of quantum evolution of complex systems using numerically efficient ODE integrators. The evolution problem thereby reduces to a statistical-mechanics problem of constructing an ensemble of the quantum characteristics and the associated Jacobi fields. After constructing the quantum characteristics, the physical observables can be found without further recourse to quantum dynamics.

A clear gap exists between the classical dynamics of a particle and its quantum dynamics. In the first case, we are dealing with a finite number of degrees of freedom. In the second case, we are dealing *per se* with field theory and an infinite number of dynamical degrees of freedom. We see that this gap is filled with the Jacobi fields of higher orders. By increasing the order of $\hbar$-expansion, the number of Jacobi fields is growing rapidly. This provides, in principle, a smooth transition from classical dynamics to quantum dynamics and from mechanics to field theory.

Quantum characteristics are useful for calculating the evolution of complex quantum systems of atoms, molecules and nuclei. The main advantage of deformation quantisation is its proximity to the classical picture of evolution in phase space. Specific quantum effects, such as coherence and non-localities, appear due to the increase in the number of dynamical degrees of freedom: Jacobi fields. The method of quantum characteristics allows for the consistent inclusion of non-localities and coherence in the transport models.

## Acknowledgment

This work was supported by Grant No. 09-02-91341 of the Russian Foundation for Basic Research.